\documentclass[aps,prl,twocolumn]{revtex4}
\usepackage{amssymb}
\usepackage{bm}
\usepackage{graphicx}
\usepackage{floatrow}
\usepackage{epstopdf}

\def\independenT#1#2{\mathrel{\setbox0\hbox{$#1#2$}
\copy0\kern-\wd0\mkern4mu\box0}} 

\begin{document}
\author{M. Nazarzahdemoafi$^1$, F. Titze$^1$, S. Machulik$^1$, C. Janowitz$^1$, Z. Galazka$^2$, R. Manzke$^1$, and M. Mulazzi$^{1,2}$}
\affiliation{Humboldt-Universit\"at zu Berlin, Institute f\"ur Physik, Newtonstrasse 15, D-12489 Berlin, Germany$^1$}
\affiliation{Leibniz Institut f\"ur Kristallz\"uchtung, Max-Born-Str.2, D-12489 Berlin, Germany$^2$}

\title{Comparative study of the electronic structures of  the In and Sn/In$_2$O$_3$(111) interfaces}

\date{\today}
\begin{abstract}
The electronic structure of the transparent semiconductor In$_2$O$_3$ has been studied by angle-resolved photoemission spectroscopy  upon deposition of metallic indium and also tin on the surface of the semiconductor. By deposition of metallic indium on In$_2$O$_3$(111) single crystals, we detected the formation of a free-electron like band of effective mass (0.38$ \pm $ 0.05) m$_0$. At low coverages, metallic In shifts the Fermi level of In$_2$O$_3$ to higher energies and a new electronic state forms at the metal/semiconductor interface. This state of two-dimensional character (2D-electron gas) is completely responsible for the electrical conduction in In$_2$O$_3$(111) at the surface region and has a band dispersion, which does not  correspond to the previously found surface accumulation layers in this material. Despite the similarity of the electronic properties of In and Sn, a larger downward banding was observed by Sn coverage, which was not accompanied by the appearance of the surface state.
\end{abstract}

\maketitle

\section{Introduction}
Being the building blocks of virtually any electronic device, from solar cells to the transistor of  integrated circuits, semiconductor materials are the cornerstone of modern technology. Despite the wide availability and the low costs of the classical materials like Si or GaAs, alternatives are being sought for applications,  involving the use of visible light (solar energy and optoelectronics) and/or high voltage drops across the semiconductor (high-power switches). Conductive oxide semiconductors are materials with suitable properties for solar cells or for high-voltage elements, since they have a wide forbidden band-gap and can, at the same time, conduct electricity. While from naive arguments one of these two properties should exclude the other, both are present in this material, challenging scientists to explain their behaviour and engineers to apply them in new devices. One of the open problems is the high conductivity even in nominally undoped materials. It was attributed to Oxygen vacancies, which could offer an alternative conduction channel, as showed in  other oxide materials like SrTiO$_3$ and its interfaces \cite{siemons}. Furthermore, accumulation layers at the surface of the oxides were reported several times in the case of thin films grown on substrates and upon adsorption of water \cite{Brinzari, Kurtz}, which allowed using In$_2$O$_3$ as a gas detector.\\
The crystals used in this work are truly bulk single-crystals grown from melt \cite{galazka1,galazka2}, and described further below. In this work, we analysed the properties of In$_2$O$_3$, a material with a band gap of around 2.7 eV \cite{scherer, king1, bourlange, irmscher}  whose conduction properties have been the subject of debate in the last few years. Numerous theoretical \cite{lany,fuchs,Karazhanov,Medvedeva,agoston,erhart} and experimental \cite{walsh,scherer,janowitz1,king2,king1,zhang,bourlange,bierwagen} works have been devoted to In$_2$O$_3$, but the microscopic origin of the conduction in this materials remains unclear. To shed light on the problem, we examined the behaviour of the electronic structure of In$_2$O$_3$ upon deposition of metallic indium and tin layers measuring the In 4d core levels and valence band. We observed that the core electron binding energies increased with respect to the clean surface case by the studied metals and a new state appearing near the Fermi level in the In$_2$O$_3$ forbidden gap at the initial stage of the growth of In. This state is located at the centre of the Brillouin zone, has parabolic dispersion, an effective mass of 0.38 $m_0$, the latter being the free electron mass, and shows no k$_\perp$ dispersion in photon-energy dependent  angle-resolved photoemission spectroscopy (ARPES) measurements. The intensity of the state increases at low In thicknesses, while the emission intensity fades away at higher coverages. Finally, a uniform metallic-like density of states is present in the whole energy gap of In$_2$O$_3$ at higher thicknesses. Metallic Sn,  possessing one more valence electron,  caused a larger downward shift of the bands in comparison to indium, but no new state was measured within the gap. We exclude that the bottom of the conduction band of In$_2$O$_3$  becomes occupied based on the lack of k$_\perp$ dispersion. The new state near the E$_F$ differs from previously reported electron accumulation layer of  the In$_2$O$_3$ because of the different effective mass value we found.
\section{Experiments}
For the present study bulk In$_2$O$_3$ single crystals were grown from the melt by the Levitation-Assisted Self-Seeding Crystal Growth Method \cite{galazka1,galazka2}. The as-grown crystals were dark and almost opaque but turned yellowish-transparent after annealing in oxygen atmosphere for at least 10 hours at 800-1000 $^\circ$C \cite{galazka1, galazka2, galazka3}. After this post-growth heat treatment, the crystals showed \textit{ n}-type semiconductor with free electron concentration of 2 $\times$10$^{17}$ cm$^{-3}$ and electron mobility of  210 cm$^2$V$^{-1}$s$^{-1}$  according to the Hall effect measurements in the Van der Pauw configuration at room temperature using In-Ga ohmic contacts \cite{galazka3}. The (111)-oriented samples of  $3\times3\times0.5$ mm$ ^{3} $ were prepared from large In$_2$O$_3$ single crystals to study  In and Sn/In$_2$O$_3$ interfaces using ARPES, with ohmic-back contacts made by using silver-epoxy glue. Clean surfaces were obtained by cleavage along the (111) planes in ultra-high vacuum (UHV). The normal emission photoelectron spectra were measured by a 5 m normal-incidence monochromated source providing photons in the  5-40 eV  energy range. Photoelectrons spectra were measured by means of a Scienta SES2002 analyzer \cite{janowitz2}. Incident photon energies of 18, 35 and 39 eV were used to probe the In$_2$O$_3$ valence band, the In 4d and Sn 4d core levels,  respectively. The measurements were performed at room temperature in a pressure better than 2 $\times$ 10$^{-10}$ mbar with an energy and angle resolution of 20 meV and 0.2$^\circ$. The kinetic-energy scale of spectra was calibrated to the Fermi level which was known from a thick-polycrystalline indium and tin film on the In$_2$O$_3$ crystal. Calibration on a polycrystalline sample yielded the same results. 
The metals ( In and Sn, purity 99.99 $\%$) were evaporated from a tungsten coil on UHV-cleaved (111) surfaces of the samples under constant heating conditions at an operating pressure of 5 $\times$ 10$^{-10}$ mbar keeping the sample at room temperature in a separate vacuum chamber.
\section{Results and discussion}

\begin{figure}[htpb]
 \includegraphics[scale=0.6]{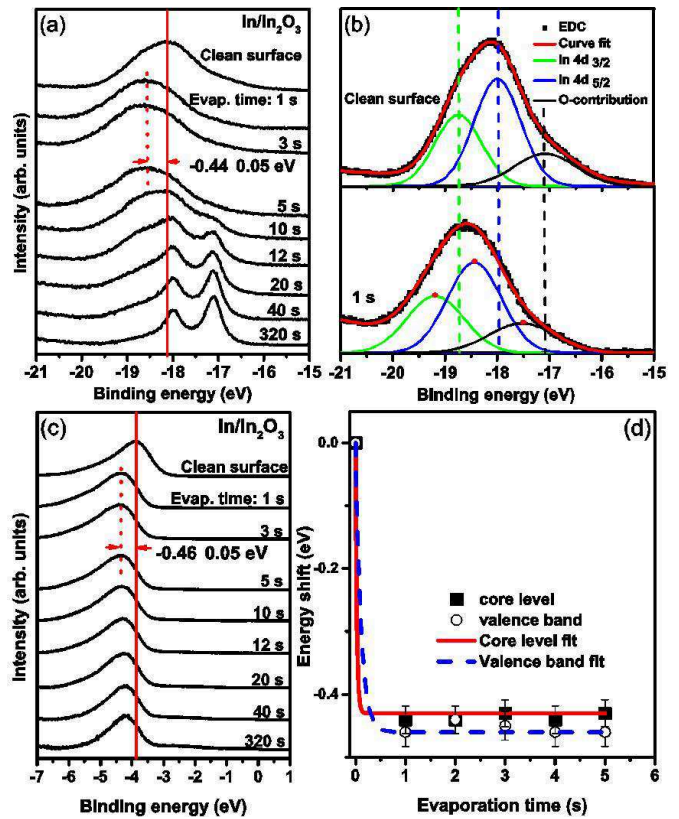}
    \caption{a)  Normal emission photoelectron spectra  of  the In-4d core level  as a function of In evaporation time at 35 eV photon energy. b) Two examples of  the fitting results of In 4d spectra for the as-cleaved surface and after 1 s of In evaporation. Each one consists of three components, the spin-orbit splitting doublet of the In 4d, and an oxygen-contribution at the lower binding energies. The red dots in the lower half show the maxima of the three fit components. c) ARPES spectra of the valence band region of In$ _{2} $O$_{3}$  versus In coverage at photon energy of 18 eV. 
d) Summary of the energy shift of the core level and valence band spectra against In-evaporation time.}
    \centering
    \label{a}
\end{figure}

The core level and valence band (VB) spectra were measured to monitor their band bending by the metal coverage. Figure \ref{a} (a) shows the core level spectra for different In coverages excited at a photon energy of 35 eV. In Fig.\ref{a} (b), two examples of the fit results for as-cleaved sample as well as after 1 s evaporation of In are shown. Similar to our previous work \cite{maryam} and also Ref. \cite{zhang}, a fit of the In 4d core level spectra was performed by three Gaussians for three components, the spin-orbit split In 4d$_{5/2}$- In 4d$_{3/2}$ doublet and the O-contribution on the lower binding energy side,  with the In 4d$_{5/2}$/In 4d$_{3/2}$ intensity ratio and spin-orbit separation was constrained to be 6/4 basing on electron counting arguments, and 0.75 eV. One can observe a clear shift of the three components of the In-4d core level to the higher binding energy after 1 s of In evaporation.

\begin{figure*}[htpb]
\floatbox[{\capbeside\thisfloatsetup{capbesideposition={right,top},capbesidewidth=4cm}}]{figure}[\FBwidth]
{\caption{a) ARPES maps of the as-cleaved and  the low In-covered samples along $\Gamma-N$ direction of the bulk Brillouin zone, corresponding to the Fermi energy region and taken at photon energy of 18 eV.  b) Parabola fit using dispersion of the nearly free-electron state near the Fermi energy region along $\Gamma-N$ direction  for a low In-covered sample (after 3 s of In evaporation). c) ARPES spectra in normal emission of near- E$_F$ region of a low-In covered sample at different photon energies from 18 to 25 eV, revealing the two dimensionality of the state within the band gap.}}
{\includegraphics[scale=0.6]{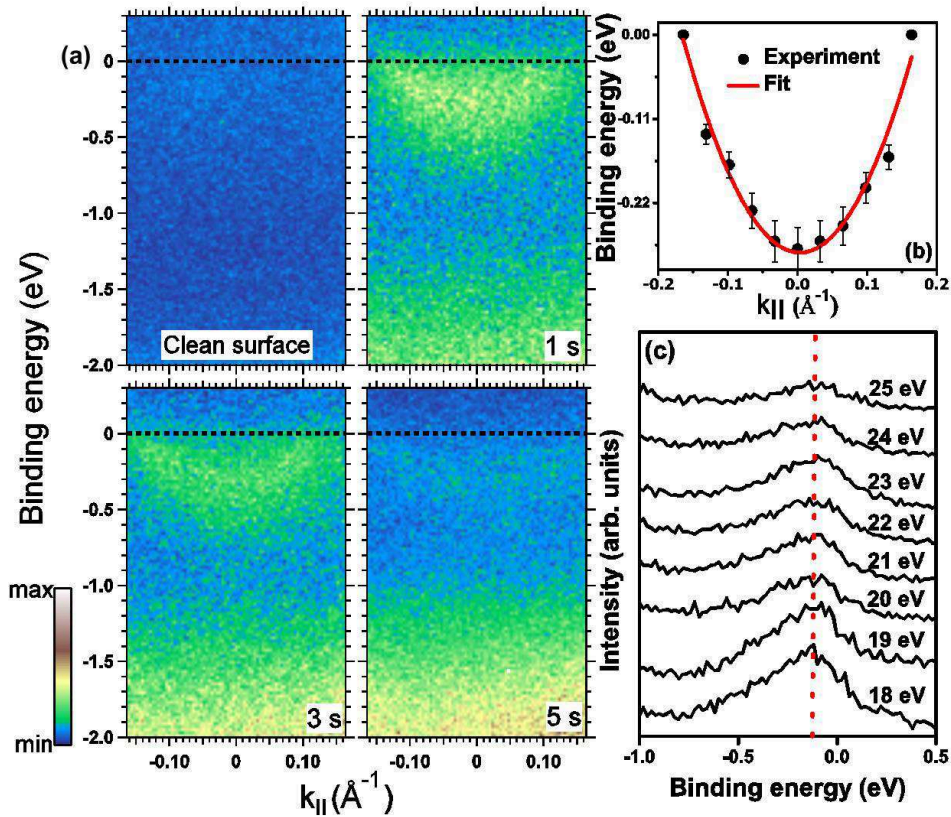}}
\end{figure*}
\begin{figure}[htpb]
 \includegraphics[scale=0.6]{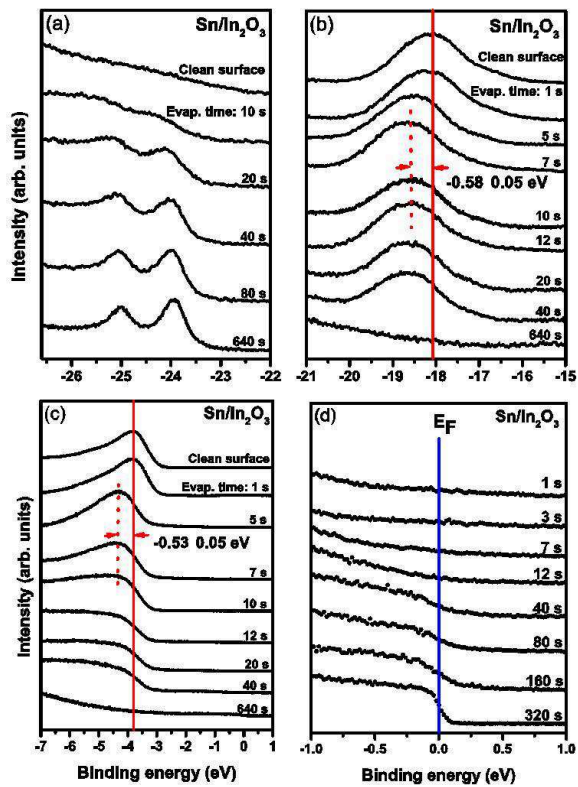}
    \caption{a) selected EDC series of the Sn 4d core level of Sn$/$In$_{2}$O$_{3}$ interfaces, two distinct spin-orbit Sn 4d$_{5/2}$-Sn 4d$_{3/2}$ doublet of the metallic Sn are observable at binding energies of 24 and 25 eV for high coverage of tin.  b), c) and d) Selected ARPES spectra series of In 4d core level, VB and near the Fermi energy regions of the clean In$_{2}$O$_{3}$(111) surface and after different evaporation times, respectively.}
    \centering
    \label{c}
\end{figure}

  In addition to the downward band bending of (0.44 $\pm $ 0.05) eV of the core level spectrum, the gradual evolution of two distinct In 4d$_{5/2}$ and In 4d$_{3/2}$ doublet of the In coverage can be noticed due to the metallic non-oxidized indium. Figures \ref{a} (c) and (d) display the ARPES spectra of the valence band  for the as-cleaved In$_2$O$_3$ samples and after various In-deposition steps, and the summary of the downward band bending of the core level and the valence band spectra, respectively. For the valence band spectra, the energy  of the most intense peak was monitored to determine the band bending. As depicted in  figures \ref{a} (c) and (d), nearly the same downward banding of the valence band spectra with In deposition was observed.
Fig. 2 (a) shows the selected photocurrent intensity maps along $\Gamma-N$ direction of the bulk Brillouin zone for the as-cleaved and the low In-covered samples near the Fermi energy recorded at $ h\nu=18$ eV.
After 1 s of In evaporation,  electron like state with parabolic dispersion appears. Its intensity increases up to 3 s evaporation time, gradually blurring and disappearing for evaporation times higher than 5 s. Figure 2 (b) shows the parabola fit of the dispersion of the state in the vicinity of the Fermi energy after 3 s In evaporation. The effective mass, binding energy of the bottom of the band, and the Fermi wave vector of a nearly free-electron state were found to be (0.38$ \pm $ 0.05) m$_0$, (0.28$ \pm $0.05) eV, and (0.16$ \pm $0.02)  \AA $^{-1}$, respectively. These values differs from those of the quantized subband states which were observed in the near-surface of In$_2$O$_3$ thin films \cite{zhang}. As indicated in the panel (c) of Fig. 2, no photon-energy dispersion of the state near the Fermi level  can be detected, revealing its  two dimensional character.  The surface charge density n$_{2D}$ is obtained to be 4.08 $\times$10$^{13}$ cm$^{-2}$ by substitution of the measured Fermi wave vector in the formula n$_{2D}=$k$_F^{2}$/2$ \pi$ \cite{piper}. This estimated surface concentration is substantially larger compared to the reported ones at classical semiconductor interfaces \cite{Ando}. In spite of the other derived parameters of this state, the obtained n$_{2D}$ is close to that for the clean surface of In$_2$O$_3$(111) single-crystalline thin film in Ref. \cite{zhang}. 
The energy difference between bulk Fermi level  and bottom of the conduction band was determined to be around 0.07 eV in the studied crystals \cite{maryam}. Considering the above-mentiond fact as well as the 2D-character of the identified state, we concluded that it is not derived from the conduction band. 
     The downward band bending of the core level and the valence band spectra in addition to the appearance of the state close to the Fermi energy in initial stage of  In- deposition, could evidence the electron doping of the surface of the In$_2$O$_3$. In fact, In might act as a surface donor  and lead to downward band bending of the levels and surface Fermi level shift to higher binding energies. To test the idea of the surface donor, we evaporated Sn, which differs from In for only one electron in the VB.
 The selected energy distribution curves (EDC) in  the Sn 4d, In 4d  core levels, VB and near E$_F$  region at Sn$/$In$_{2}$O$_{3}$ contacts are shown in figures \ref{c} (a) to (d), respectively. 
As shown by spin-orbit split Sn 4d doublet upon deposition in panel (a), the metallic Sn covers the sample with no sign of formation of  Sn-O compounds, as expected. From Fig. \ref{c} (b) and (c), one can see that  in comparison to In, Sn causes a larger downward band bending of (-0.58 $\pm $ 0.05) eV. This larger shift to higher binding energies can be attributed to stronger donor character of Sn with respect to  In.  Fig. \ref{c} (b) reveals that In 4d-core level spectra line shape does not change and it is different from that  of  Sn- doped  In$_{2}$O$_{3}$ (ITO) by Sn deposition, Ref. \cite{zhang}. Therefore, ITO formation during the growth of Sn on In$_{2}$O$_{3}$ single crystals can be discarded. Sn coverage results in the suppression of the In 4d states. The valence band line-shape changes upon tin deposition. The VB of Sn is nearly located in the same energy region. In contrast to the In$/$In$_{2}$O$_{3}$, no state within the band gap could be identified in Sn-covered samples, as shown in Fig. \ref{c} (d). 
 
The  work functions of thick layer of  In and Sn on the In$_{2}$O$_{3}$ samples were measured \textit{ in situ} and determined to be (4.02 $\pm $ 0.05) eV and  (4.31 $\pm $ 0.05) eV, respectively. By applying these values and also the \textit{ in situ}-deduced electron affinity of the as-cleaved studied crystals (4.18 $\pm $ 0.06) eV, the Schottky-Mott rule \cite{sze} predicts a barrier height of  (-0.16 $\pm $ 0.11) eV  for In$/$In$_{2}$O$_{3}$ interfaces and (0.13 $\pm $ 0.11) eV  for Sn$/$In$_{2}$O$_{3}$ contacts. These results evidenced the disagreement of the present experimental barrier heights and the predicted ones from this rule, especially for the case of Sn$/$In$_{2}$O$_{3}$ contacts.
Similar to our previous work \cite{maryam}  and also Ref. \cite{wenckstern}, the laterally-homogeneous barrier height $\phi^{hom} _{B}$, was determined for both contacts within the metal-induced gap states (MIGS) based models \cite {moench}.
By employing the electronegativity of the studied metals in Miedema unit \cite{miedema} X$_{In}$= 3.90 eV and  X$_{Sn}$=4.15 eV,  the barrier heights were derived to be -1.12 eV and -1.02 for  In and Sn$/$In$_{2}$O$_{3}$ contacts, respectively. These results also confirm ohmic character of the contacts, but the obtained barrier heights via this model are incompatible with experimental ones. The difference is not as large as former reported metal-In$_{2}$O$_{3}$ contacts \cite{maryam, wenckstern}. However, it suggests the improvement of  the MIGS model and especially charge neutrality level (CNL) concept  for In$_{2}$O$_{3}$  which might be the origin of this discrepancy.

\section{Conclusion} 
To conclude, by ARPES we showed that despite the similarity between In and Sn, the Sn and In$/$In$_{2}$O$_{3}$ interfaces behave differently. These metals act as surface donors generating downward band bending. The obtained barrier height was larger for the case of  Sn$/$In$_{2}$O$_{3}$ consistent with its valence-electron configuration. At initial stage of In growth, the downward band bending is accompanied by an emerging 2D-electron gas, which vanishes at higher coverages. The deduced information from the band dispersion of the surface state differs from the frequently-reported SEAL in In$_{2}$O$_{3}$. In contrast to In growth, no surface state was identified upon Sn deposition. The dissimilarities of the electronic structures of  the In and Sn-In$_{2}$O$_{3}$ interfaces despite the comparable structure of these metals, in addition to disagreement of the experimental results and the Schottky-Mott rule prediction indicate the complexity of contact formation mechanism in these interfaces, whose full explanation should be addressed by ab initio theoretical calculations.

\section{Acknowledgements}
This work was conducted at BESSY II. We are grateful to K. Irmscher for the electrical characterization and the staff of BESSY for their technical support. This work was financially supported by the DFG (German Research Foundation) under Project No. MA2371/8-1.

\end{document}